# Review on Titania Nanopowder- Processing and Applications

*T.Theivasanthi, International Research Center, Kalasalingam University, Krishnankoil– 626126, India. Phone: +91-4563-258084. Mobile: +91-9524818862, 9344643384. E-mail: ttheivasanthi@gmail.com*

## Abstract

This review work focuses on the milling process of titania ($TiO_2$) nanopowder. $TiO_2$ is a naturally available material. It is an inorganic semi-conducting oxide of titanium metal. Titanium oxide, titanium dioxide, and titanium (IV) oxide are the other names. $TiO_2$ nanopowder is one of the well-known synthetic materials in the modern nanotechnology field. It has many interesting, lucrative and salient properties. These characters are attracting the researchers towards to exploit this material in variety of applications from energy to environmental. It is a highly sensitive and excellent photo-catalyst material. So, it can be used in gas sensors, in water splitting to generate hydrogen, and in waste water treatment to degrade the organic pollutants. It is soluble in ethanol, ethylene glycol and chloroform. $TiO_2$ nanopowder is prepared by many methods such as sol-gel method, chemical vapour deposition, hydro-thermal, micro-emulsion etc. However, low cost and large scale synthesis of nanoparticles for commercial applications are the major challenges in the present day scenario. Mechanical milling is a simple method to prepare nanopowder where milling time plays very important role. The milling time may vary from 0 to 60 hour. The crystallinity, phase, size, shape and surface features are some of the factors affected by the milling process. For the preparation of $TiO_2$ nanomaterial, the mechanical milling is more advantageous than other methods. It is a low cost, non toxic method which offers high chemical stability. The finely grinded $TiO_2$ nano powder offers shiny surface and brilliant colours. Milling of $TiO_2$ is done for size reduction. Doping of $TiO_2$ is done to improve the efficiencies and to minimize drawbacks.

## 1. Introduction

Titania ($TiO_2$) is available in three phases namely anatase, rutile and brookite. The phase and particle size affect the physical properties of the material. When comparing to brookite and rutile crystalline structures, anatase crystalline structure appears to be more active. $TiO_2$ is one of the main white pigments widely used in paints, printing inks, plastics, cosmetics and pharmaceuticals [1-3]. The paint industry is in the leading position and it utilizes the 60 % of global $TiO_2$ pigment production [4]. Pigment $TiO_2$ with crystal size around 200 nm has many excellent qualities due to its high refractive index and significant scattering properties. Its rutile crystalline form has scattering efficiency higher than anatase crystalline [5] because it has higher refractive index *i.e.* 2.74 and anatase has 2.54 only [6].

$TiO_2$ is n-type semiconductor with 3.02 eV bandgap which is useful for solar cells mainly to enhance their efficiency. In future, it will be exploited to absorb UV, visible and infrared regions of sunlight with reduced bandgap. Metal powders doping and accelerating the solid state reactions at very high temperatures reduce the bandgap. The dispersion factor of the pigments depends on the refractive index and the particles size distribution. $TiO_2$ has highest refractive index of all the white pigments. It has refractive index higher than diamond and many other known materials. It has an extraordinary capacity to disperse light in the visible part of the light spectrum. It is mixed with paints, plastics, paper and cosmetics as white pigment. It is also added with car paint and decorative coatings for furniture.

Photo-catalytic materials decontaminate or degrade the various organic and inorganic pollutants. Their performance has closely connected with their properties such as Phase, crystallinity, crystal size and surface properties. These properties are affected by preparative conditions and techniques adopted during their productions. Noble metal or non-metal doping on the $TiO_2$ surface enhances the efficiency of $TiO_2$.

Bio- chemical stability, low cost, non-toxic and long stable life are some promoting factors to utilize $TiO_2$ as a photo-catalyst. But, it absorbs only a small fraction from the visible region of sunlight. This property limits its photo-catalytic efficiency and concomitant photo-catalytic ability. Decreasing its band gap energy range from 3.02 eV or preventing electron-hole pair ($e^-$ / $h^+$) recombination by doping process will improve its photo-catalytic related activities. Such doped $TiO_2$ with low band gap will be applied to absorb both UV and visible light.

Metal oxide like $TiO_2$ has ability to sense the LPG. It has potentials in gas detection due to low cost, stable phase, desirable sensitivity, high chemical stability and non-toxicity. Commercially, it is used as lambda sensor in exhausted pipes. Materials tend to trap electrons from semiconductor are easily absorbed by $TiO_2$. The optical band gap value of $TiO_2$ is significant for gas sensing point of view.

$TiO_2$ is grinded in ball mill for size reduction as well as for doping purposes. $TiO_2$ powder is produced with fine particles (from macro to nm size) depends on the milling conditions such as speed, time etc. The nano level particles are achieved by milling method. They accelerate solid-state reactions at high temperatures. Possibilities are more to utilize $TiO_2$ nanomaterial prepared by ball milling in various applications. UV and photo-catalytic activities of $TiO_2$ increase after the ball milling process of $TiO_2$ with doping materials. On the other hand surface area decreases gradually while increasing the doping material quantity. Surface area is opposite to the milling speed. The milling process reduces mean particle size.

Variations in rotating speed, grinding time, balls to grinding materials ratio and concentrations of the materials are some of the factors implemented to get the desired effects. Milling of $TiO_2$ with variations in these factors has been reported in literatures. Already, such procedures have been followed by many researchers. Also, currently is in practice. This milling method is used to grind either $TiO_2$ separately or $TiO_2$ as doping material with other materials or other doping materials with $TiO_2$. Improving the efficiency of $TiO_2$, to create changes in the properties of $TiO_2$ and other materials, to create new properties in these materials and to create new composites materials are some of the expected yields from this milling. These new or improved or modified properties can be exploited in many applications and in various utilizations.

## 2. Titania Nanomaterial

Zhang *et al.* report, the specific surface area and surface to volume ratio increase dramatically as the size of materials decreases. The high surface area (SA) of $TiO_2$ nanoparticles facilitates the reaction / interaction between $TiO_2$ based devices and the interacting media, which mainly occurs on the surface or at the interface and strongly depends on the surface area of the material [7]. Addition of $TiO_2$ nanoparticles with cement based materials increase the mechanical properties like compressive and flexural strength. It is also used for the self-cleaning coating and sterilizing properties in concrete. It blocks UV light and provides photo-catalytic properties to traditional cements.

Titania nanoparticles are incorporated in paints, cements and windows for its sterilizing properties and to block UV light. $TiO_2$ breaks down organic pollutants, volatile organic compounds and bacterial membrane through powerful catalyst reactions which reduces airborne pollutants of outer surfaces. Also, it has hydrophilic property which gives self-cleaning properties to surface on which it is applied [8]. $TiO_2$ has been used to tailor photo-catalytic properties of traditional cements [9].

Studies on mechanical properties of cement based materials with nanomaterials of $SiO_2$, $TiO_2$ and $Fe_2O_3$ confirms the increasing of compressive and flexural strength. Some *major developments:* Well-dispersed nano-particles increase the viscosity of the liquid phase helping to suspend the cement grains; improving the segregation resistance and workability of the system; nano-particles fill the voids between cement grains, resulting in the immobilization of "free" water; act as centers of crystallization of cement hydrates, therefore accelerating the hydration; crack arrest and interlocking effects between the slip planes provided by nano-particles improve the toughness, shear, tensile and flexural strength of cement based materials; also improves the bond between aggregates and cement paste [10].

## 3. Milling- Size Reduction & Fineness

The milling is a simple method for reducing the particle size with various levels *i.e.* from macro to nano level. Ball milling is one of the effective mechanical milling processes and the milling time plays very important role. During this process, wet or dry milling condition is adopted. The decreasing of particles size depends on the powder-to-ball weight ratio.

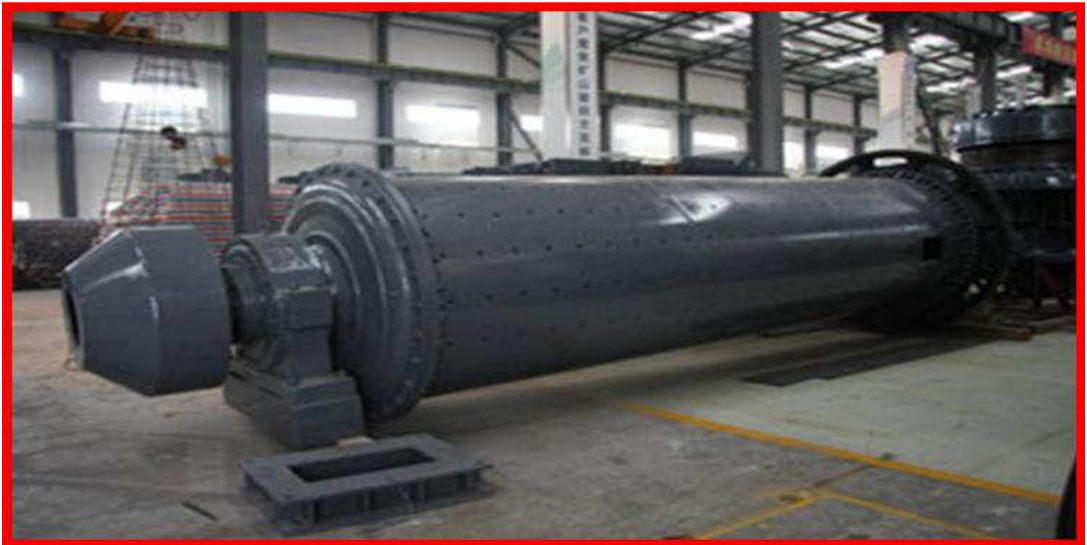

**Fig.1.** Commercial Ball Mill image. Figure is adapted from [11].

Ball mill image is shown in Fig.1. Ball mills are the main equipment in grinding industry and suitable to grind high hardness material without pollution. They are in two varieties *i.e.* wet and dry mills. They are used primary for single stage fine grinding and regrinding. Also, they are used as the second stage in two stage grinding circuits. Materials such as cement, silicates, building / fire-proof material, fertilizers, non-ferrous metal, glass, ceramics etc. can be grinded. Shape of the final product is circular. Its fineness can be tuned by adjusting diameter of the balls [11].

In ball milling, known macro-crystalline structures are broken down into nano-crystalline structures, but the original integrity of the material is retained. However, the nanoparticles can reform into new a material, which involves breaking the original crystalline bonds. Among these methods high energy milling has the advantages of being simple, relatively inexpensive, and applicable to any class of materials which can be easily scaled up to large quantities [12].

A stirred media mill is a vertical or horizontal cylinder which is loaded with grinding beads (size range from 0.05 mm to 3 mm) made from steel, glass, ceramics or plastics. Also, it is filled with 60 to 85 % suspension containing the particles for grinding. A rotating stirrer sets the beads and the suspension in motion whereupon the particles are stressed by compression, impact and shearing brought about by collision of the beads, causing size reduction [13].

Grinding in a wet stirred media mill is the last step for the distribution of submicron or nanoparticles before added to an application. Since stirred media milling is an energy-intensive process and energy efficiency should be optimized. This can be done by determining the optimum operational parameters for the mill and using the highest possible solids concentration. The solids concentration can be increased by controlling particle-particle interactions with stabilizing chemicals like polymers [14].

## 4. Milling of Titania

### 4.1. TiO$_2$ Nanopowder- Grinding

TiO$_2$ powder size is reduced by dry grinding of TiO$_2$ powder [15] which is explicated in this sub-section. Physical grinding method is adopted to reduce the size of TiO$_2$ powder to TiO$_2$ nanopowder. TiO$_2$ powder is grinded for 15 minutes in a high speed rotating grinding machine. At the end, TiO$_2$ nanopowder is separated.

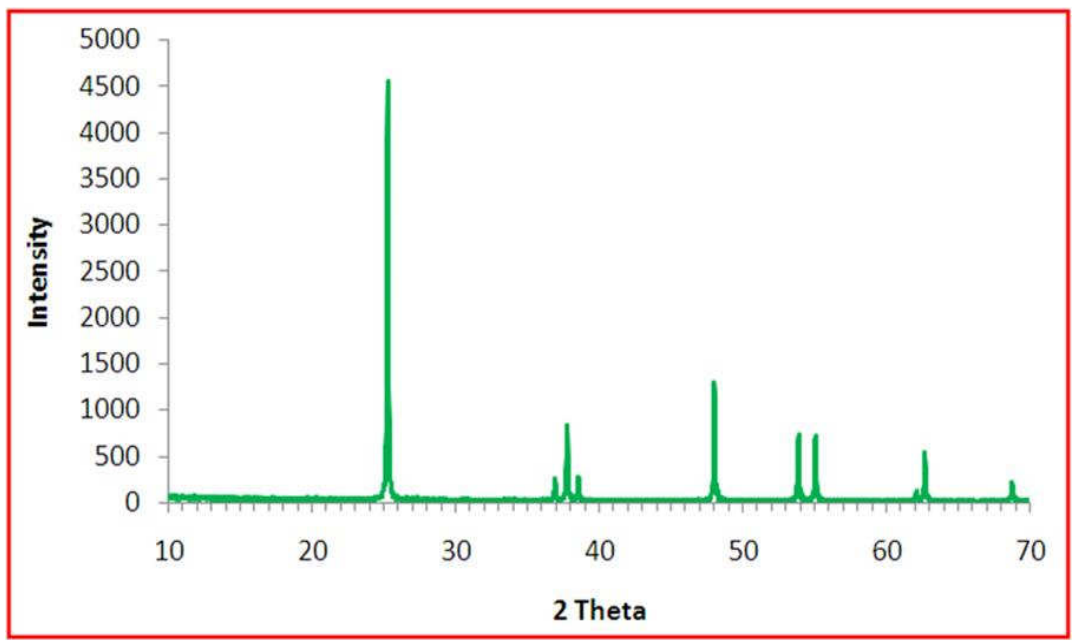

**Fig.2.** XRD pattern of TiO$_2$ nanoparticles. Fig. is adapted from [15].

The XRD pattern of the synthesized Titania nanoparticles is shown in Fig.2. Particle size, specific surface area (SSA), dislocation density (δ) and morphology index (MI) are calculated from the XRD studies of $TiO_2$ nanopowder. Particle size (74 nm after instrumental broadening) is estimated by using *Debye-Scherer formula* (1). Inter-planar spacing between atoms (d-spacing) is calculated using *Bragg's Law* (2).

$$D = \frac{0.9\lambda}{\beta cos\theta} \quad\quad\quad\quad\quad\quad\quad\quad\quad\quad\quad\quad\quad\quad\quad\quad\quad\quad\quad\quad (1)$$

$$2dsin\theta = n\lambda \quad\quad\quad\quad\quad\quad\quad\quad\quad\quad\quad\quad\quad\quad\quad\quad (2)$$

Where, λ is wave length of X-Ray (0.1540 nm), β is FWHM (full width at half maximum), θ is diffraction angle, d is d-spacing and D is particle diameter size.

Mathematically, SSA can be calculated using formulas (3 & 4) and both formulas yield same result (19.16 $m^2g^{-1}$).

$$SSA = \frac{SA_{part}}{V_{part} * density} \quad\quad\quad\quad\quad\quad\quad\quad\quad\quad\quad\quad (3)$$

$$S = 6 * 10^3 / D_p \rho \quad\quad\quad\quad\quad\quad\quad\quad\quad\quad\quad\quad\quad\quad (4)$$

Where SSA & S are the specific surface area, $V_{part}$ is particle volume and $SA_{part}$ is Surface Area, $D_p$ is the size (spherical shaped), and ρ is the density of $TiO_2$ (4.23 $g.cm^{-3}$).

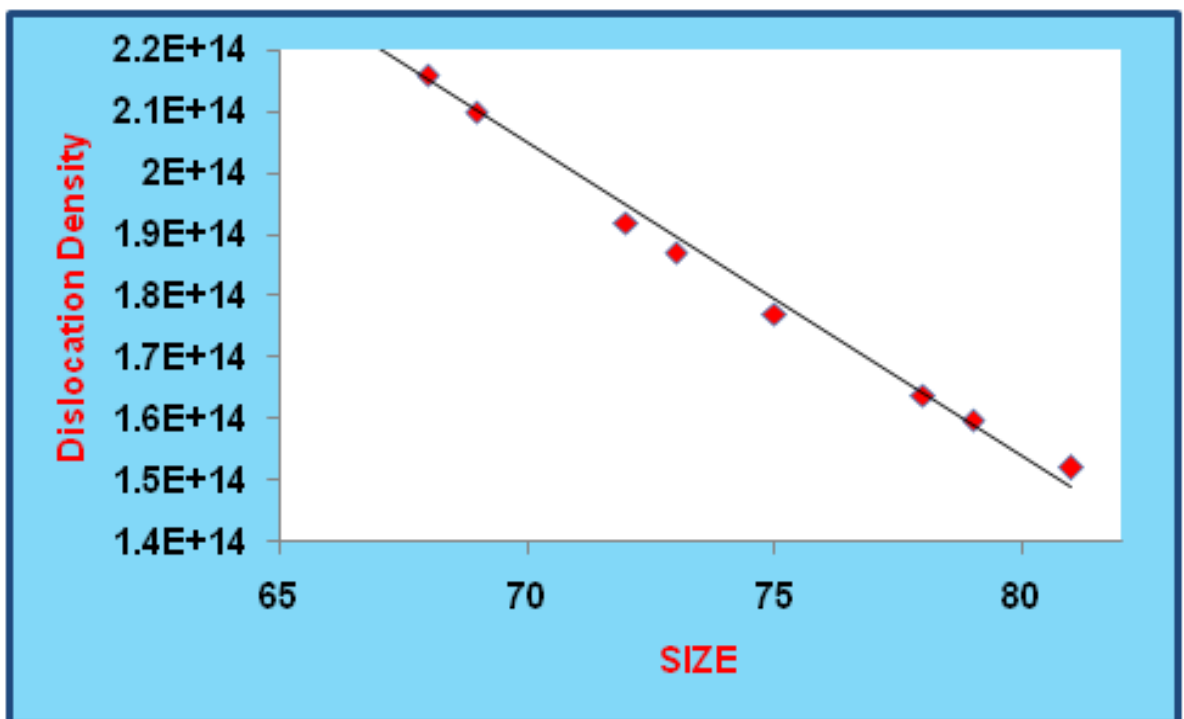

**Fig.3.** Particle Size Vs Dislocation Density of $TiO_2$ Nanoparticles. Fig. is adapted from [15].

The dislocation density (δ) of the sample is inversely proportional to particle size. Larger δ implies a larger hardness. The δ of the sample is 1.82 x $10^{14}$ $m^2$. It implies that the prepared $TiO_2$ nanoparticles have strength and hardness more than their bulk ($TiO_2$) counterpart. The δ of the sample is determined using expressions (5 & 6). The results from both formulas are approximately same.

$$\boldsymbol{\delta} = \frac{15\beta cos\theta}{4aD} \dots\dots\dots\dots\dots\dots\dots\dots\dots\dots\dots\dots\dots\dots\dots\dots\dots\dots(5)$$

$$\boldsymbol{\delta} = 1/D^2 \dots\dots\dots\dots\dots\dots\dots\dots\dots\dots\dots\dots\dots\dots\dots\dots\dots\dots(6)$$

Where, δ-dislocation density, β- diffraction broadening - measured at half of its maximum intensity (radian), θ- diffraction angle (degree), α-lattice constant (nm) and D-particle size (nm). The correlation between δ and particle size is shown in Fig.3.

MI is calculated to explore the relationship between particle morphology, size and specific surface area of $TiO_2$ nanopowder. It is obtained using expression (7)

$$\boldsymbol{M.I} = \frac{FWHM_h}{FWHM_h + FWHM_p} \dots\dots\dots\dots\dots\dots\dots\dots\dots\dots\dots\dots\dots\dots(7)$$

Where, $FWHM_h$ is highest FWHM value obtained from peaks and $FWHM_p$ is particular peak's FWHM for which M.I. is to be calculated. It is observed that MI is directly proportional to particle size and inversely proportional to specific surface area. The observed results of the MI confirm the uniformity and fineness of the prepared nanoparticles.

### 4.2. $TiO_2$ Nanopowder- Mechano-chemical

Anuradha *et al.* compare the efficiency of three different synthetic routes for the production of nanostructured $TiO_2$, *i.e.* sol-gel method involving template, mechano-chemical and combustion synthesis. The microstructures of nanocrystalline $TiO_2$ produced by the above three methods are analysed. In mechano-chemical synthesis, stoichiometric amounts of titanium and cupric oxide are mechanically milled in a planetary ball mill. Stainless steel vial with wear resistant stainless steel balls are used in this process. Combustion ball milling/ mechano-chemical reaction between titanium and cupric oxide in air resulted in the successful formation of nanocrystalline $TiO_2$. After 4 h milling time, the average particle size is < 50 nm which is further reduced to ~ 20 nm after 12 h [16].

### 4.3. $TiO_2$ – Aqueous Suspension

Sato *et al.* analyse the dispersion of $TiO_2$ particles in aqueous medium. Different mechanical dispersion methods such as ultrasonic irradiation, milling with 5 mm diameter balls, and milling with 50 μm beads are adopted to analyse $TiO_2$ nano and sub-μm level particles in aqueous suspension. Poly-acrylic acid (1200 to 30000 g.$mol^{-1}$ molecular weights range) is used as dispersing agent. There are no appreciable changes in the viscosities and aggregate properties of the sub-micrometer powder suspensions during ultrasonic irradiation and 5 mm ball milling dispersion. Aggregates of nanopowder suspension are dispersed well by 50 μm bead milling like ultrasonic irradiation method but the only difference is low level solid content. It is concluded that the ultrasonic dispersion method produces highly concentrated with good dispersed nanoparticle suspensions [17].

## 5. Milling of Titania with Doping Materials

### 5.1. Doping of $TiO_2$ and Energy band gap

$TiO_2$ has various high efficiency properties and some drawbacks. Many endeavors have been attempted by researchers to improve the efficiencies as well as to overcome the drawbacks. Reducing the band gap energy helps in providing solutions for some drawbacks and for improving catalytic activity. Reducing the size improves the surface area and mechanical strength but it increases the bandgap. Some researchers observe that doping in $TiO_2$ with non-metal material such as nitrogen or sulphur is one of the good solutions which reduce the band gap. They adopt such doping practice using ball milling.

$TiO_2$ is considered as a multipurpose material due to wide band gap energy, low cost, non-toxic, reusable, chemical stability under UV light and high resistant against chemical corrosion. Also, this material has high efficiencies in chemical, mechanical and optical properties which lead to various applications such as dye sensitized solar cells, self-cleaning, photo-catalysts and reducing pollution in the environment [18-20].

But, it has some drawbacks *i.e.* fast recombination rate of electron-hole pair and light activation under visible light which limit the activities. Reducing the band gap energy overcomes these problems. Non-metal doping into $TiO_2$ particle is an effective method to reduce band gap and to enhance catalytic activity [21-22]. Ball-milling process is one of versatile methods which reduce particle size from macroscale to nanoscale. The particle size is reduced by rupturing ability of the grinding media. Heat treatment enhances the photo-catalytic reaction and improves the surface area by removing impurities [23]

### 5.2. $NH_3$ Doped $TiO_2$ – Ball milling

N-doped $TiO_2$ is produced in ball mill by Techitdheera *et al*. [24]. It is explicated in this sub-section. Commercial $TiO_2$ anatase powder is dispersed in ammonia solution in ball milling process to prepare N-doped $TiO_2$. This process is done in a horizontal planetary ball mill contains zirconia balls with a diameter of 2 mm and 5 mm (Ball volume is 55 %) at room temperature for three different milling periods *i.e.* 24 h, 36 h and 48 h. After completion of the milling process, the powder is annealed under nitrogen atmosphere at two different temperatures *i.e.* 200 °C and 300 °C for 1 h with heating rate 1 °C per minute. The active surface areas of the end product are measured by Brunauer Emmet Teller (BET) method. It is observed that the increasing the milling period removes the contamination of $H_2Ti_3O_7$ phase (hydrogen titanate) in $TiO_2$ powder and improves surface area. Also, milling time is the main parameter for the morphologies of N-$TiO_2$. $NH_3$ solution is acting as good nitrogen source for the incorporation of nitrogen in Titania. XRD of patterns of commercial $TiO_2$ powder, milled $TiO_2$ powder (after 48 h milling) after annealing at two different temperatures (200 °C and 300 °C) are compared to investigate the milling and annealing effects.

## 5.3. Red Phosphorous Doped $TiO_2$

RP-$TiO_2$ nano-hybrid photo-catalyst is prepared using milling process [25]. It is explicated in this sub-section. The catalytic efficiency of $TiO_2$ is low due its wide band gap and high recombination rate. Several attempts like metal and non-metal doping, anchoring of noble metals, textural designing and surface defect have been made to enhance the photo-catalytic activity. One of such attempts *i.e.* combines the fascinating properties of the RP (wide visible light absorption and low band gap) with $TiO_2$. It improves the catalytic efficiency of $TiO_2$.

Ansari *et al*. report that the elemental red phosphorus (RP) doping in $TiO_2$ improves the visible light absorption ability and photo-catalytic activity of $TiO_2$. The doping process is done using high energy ball milling. Optimal RP loading and milling time are confirmed from the report. The use of earth abundant and inexpensive RP instead of plasmonic metals is useful in solar cells for the efficient energy conversion from visible light.

Commercial RP powder and $TiO_2$ nanoparticles are grinded in horizontal oscillatory ball mill to synthesize the RP-$TiO_2$ nano-hybrid photo-catalyst. They are milled at 250 rpm for different time periods, *i.e.* 0, 12 and 24 hour. The weight ratio of the precursors is 50:50. The photo-catalytic activity of the end product is compared with RP-$TiO_2$ nano-hybrid powder prepared in two different weight ratios. For this purpose, nano-$TiO_2$ with 10 % of RP and nano-$TiO_2$ with 20 % of RP are prepared using the above mentioned method. A control photo-catalyst is also prepared by hand grinding the same composition of nano-$TiO_2$ and RP (RP-$TiO_2$-mix), as mentioned above.

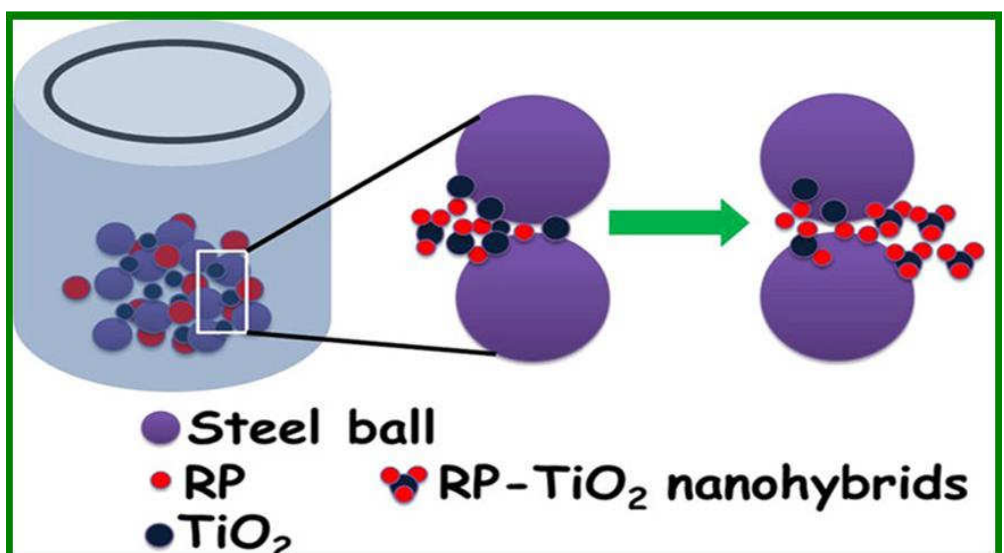

**Fig.4.** High energy ball milling of RP-$TiO_2$ nano-hybrids. Fig. is adapted from [25].

Fig.4. presents the mechanism RP doping in $TiO_2$. Ball mill effectively grinds large particles into the nano-regime through the high energy forces developed by colliding high speed stainless steel balls with the reactant media and the cylindrical shell. As a result of the repetitive fracturing and welding process leads to solid-state reaction and the formation of a nano-hybrid photo-catalyst.

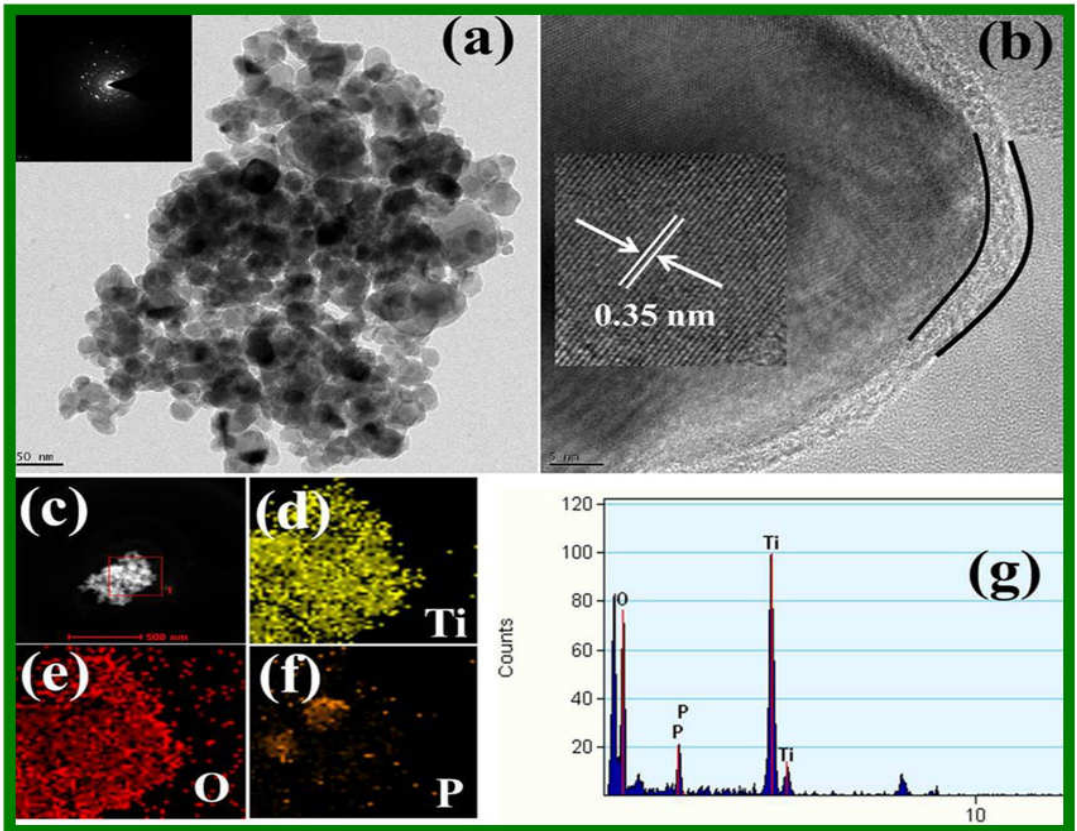

**Figure 5.** (**a**) TEM image with SAED pattern inset, (**b**) HR-TEM image with lattice fringe inset, (**c–f**) scanning transmission electron microscopy elemental mapping, and (**g**) EDX of the 12h milled RP-TiO$_2$ nano-hybrid. Figure is adapted from [25].

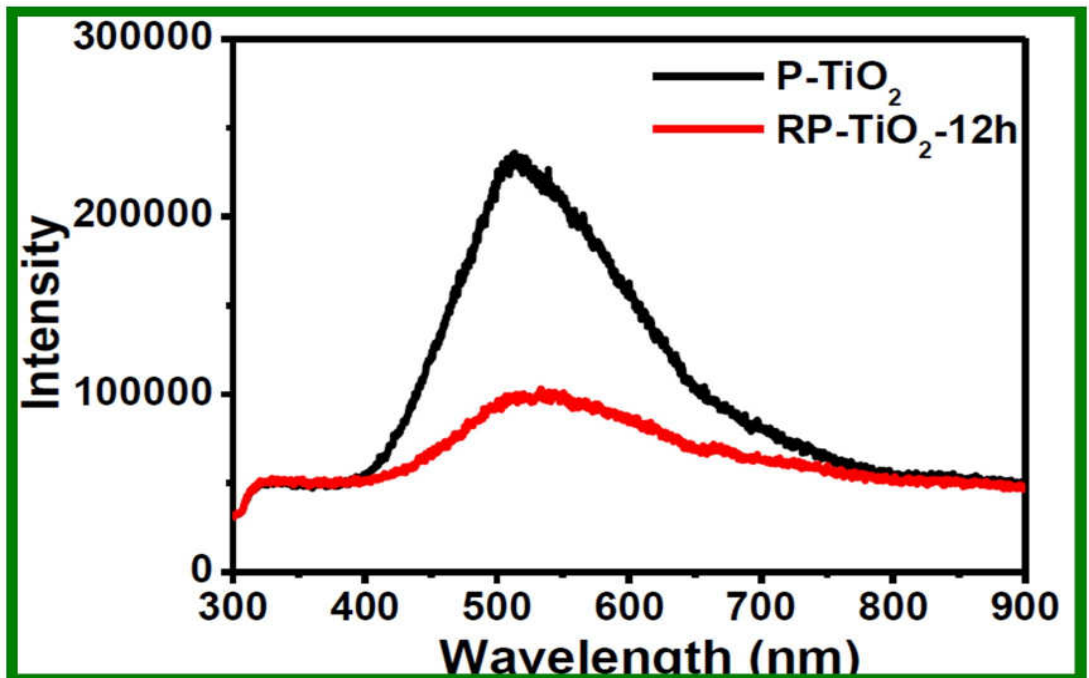

**Figure 6.** PL spectra of P-TiO2 and 12 h milled RP-TiO2. Figure is adapted from [25].

HR-TEM, STEM images and EDX of 12 h milled RP-$TiO_2$ nano-hybrid are shown in Fig. 5. Photoluminescence (PL) spectra of 12 h milled RP-$TiO_2$ and P-$TiO_2$ (pure $TiO_2$) are illustrated in Fig. 6. The strong and lower emission intensity in the PL spectra indicates the high and low electron-hole recombination rate respectively. The low recombination rate enhances the photo-catalytic activity of the material. The PL emission intensity of 12 h milled RP-$TiO_2$ is very low when compared to the intensity of P-$TiO_2$. It is due to the lower recombination rate of the photo-generated electrons and holes. The presence of RP on the $TiO_2$ surface improves the separation of photo-induced electrons and holes which lead to the suppressed recombination in the 12 h milled RP-$TiO_2$ nano-hybrid. It is cleared that 12 h milled RP-$TiO_2$ possesses high photo-catalytic activity.

### 5.4. $TiO_2$ - Photo-catalyst

Zhou *et al.* has adopted a facile ball milling method for the preparation of $C_3N_4$/$TiO_2$ nanoparticles [26]. It is explicated in this sub-section. This hybrid material has shown enhanced photo-catalytic performance due to high separation and migration efficiency of electron-hole pairs. The mechano-chemical process has dispersed the $C_3N_4$ (carbon nitride) on the surface of $TiO_2$ particles. Also, it has formed layers (single layer hybrid structure & multi-layer core-shell structure) of g-$C_3N_4$ (Graphitic carbon nitride). Variations in milling period, milling speed and $C_3N_4$ loading quantity have been adopted during milling process to analyze the properties of the end product. It has been observed that the improved photo-catalytic activities and photocurrent response of $C_3N_4$/$TiO_2$ under visible and UV light irradiation.

Also, the effect of ball milling speed on the surface area and pore size of 3 %-$C_3N_4$/$TiO_2$ sample. Both, specific surface area ($S_{BET}$) and pore size of $C_3N_4$/$TiO_2$ depend on the secondary processes like aggregation and agglomeration. They are inversely proportional to milling speed. They decrease while milling speed increases. The reverse direction of surface area with milling speed indicates the aggregation of particles. The value of $S_{BET}$ decreases gradually when $C_3N_4$ loading increases.

The photo-catalytic activity is influenced strongly by ball milling time. The optimum milling time is 3 h, beyond that the photo-catalytic activity decreases gradually with increasing milling time. It is due to increased activity sites and crystal lattice distortion of $TiO_2$.

It is observed that, after ball milling for 3 h, the photo-catalytic activity of 3 %-$C_3N_4$/$TiO_2$ increases 1.2 times more than the pure $TiO_2$. It indicates that the enhanced UV activity and the increased number of photo-catalytic active sites of $C_3N_4$/$TiO_2$ hybrid structure. After the optimum milling time of 3h, fresh surfaces are formed by ball milling agglomerate.

## 6. Milling of Materials with $TiO_2$ as Dopant

### 6.1. $TiO_2$ – Bio-Applications

Titania powder is a bio-compatible and non-toxic material. This biological benign material is utilized in various bio-medical applications. It is reinforced with bioactive material to improve the strength of medical implant applications. It is applied as anti-microbial coating in health care systems to control infections.

Geetha *et al.* report about titania nanopaint based on $TiO_2$ nanoparticles embedded alkyd resin matrix. Ball milling process is used for embedding. The experimental results showed that the $TiO_2$ nanopaint coated surfaces inhibited about 85 %, 90 % and 76 % of microbial population against *Pseudomonas aeruginosa, Staphylococcus aureus* and MRSA (methicillin resistant *Staphylococcus aureus*) respectively. These results confirm that the $TiO_2$ nanopaint antibacterial coating is very useful in the control and prevention of healthcare associated infections [27].

**6.2. $TiO_2$- Reinforcing in HAp**

Bone is a composite material consisting hydroxyapatite (HAp) crystals as a main phase. These Hap crystals are embedded in bio-organic matrix. The immune system of the human body rejects all foreign materials except chemically same to body tissues [28]. HAp is a bioactive material, chemically similar to the bones of mammals. But it does not have enough mechanical strength for long term applications [29]. To increase the mechanical property of HAp powder, $TiO_2$ powder is reinforced in it.

Conversion of $TiO_2$ powder into $TiO_2$ nano-powder and reinforcing them in HAp powder are done by Mishra *et al.* using ball mill [30]. These processes are explicated in this sub-section. Reinforcement of $TiO_2$ powder and $TiO_2$ nano-powder into HAp powder makes $TiO_2$/HAp composite and $TiO_2$/HAp nanocomposite respectively. $TiO_2$ nanoparticles are prepared by mechanical milling method in two different conditions (dry and wet milling) with different milling periods (20 h and 35 h).

For both milling, commercially available micron sized $TiO_2$ powders are grinded in a planetary ball mill (has tungsten carbide balls and vials). Weight ratio of the ball to the powder is 10:1 and the rotating speed for milling is 300 rpm. Dry milling is performed without liquid milling medium. Acetone is used as the medium in wet milling process. It is found that the dry Mechanical Milling is better than wet Mechanical milling process in terms of Particle size (12 nm) and Hardness. It provides very smaller size particles when comparing to wet milling. Milling time is inversely proportional to particle size.

Both HAp based composite and nanocomposite are prepared separately using low energy ball milling technique. It provides a novel way for the dispersion of Titania particles into HAp matrix. $TiO_2$ coarse powder sample 10 % w is dispersed in HAp powder (99 % purity with an average particle size ~20 μm) with toluene liquid medium are grinded in a ball mill to obtain the $TiO_2$/HAp composite. The ball mill has tungsten carbide vials and balls. The milling process is done at 300 rpm speed for 1 h duration. Likewise, 5 % w $TiO_2$ nanoparticles are separately dispersed in HAp powder to make $TiO_2$/HAp nanocomposite.

Both the composites are harder than the pure HAp. $TiO_2$ nanoparticles reinforced composite shows hardness higher than the $TiO_2$ coarse powder reinforced composite. Also, it shows higher hardness at lower concentrations of filler materials (5 % $TiO_2$ nanoparticles in HAp and 10 % coarse $TiO_2$ particles in HAp.). This is due to reduced dislocation movement by reinforcements and enhanced diffusivity of nanoparticles.

TEM analysis shows nanoparticles in the $TiO_2$/HAp nanocomposite. SEM analysis shows lower porosities in the $TiO_2$/HAp nanocomposite when comparing to $TiO_2$ coarse particles reinforced counterpart ($TiO_2$/HAp composite). Micro hardness analysis shows the hardness improvement in $TiO_2$/HAp nanocomposites.

### 6.3. $TiO_2$ and Thermoelectric

Efficient thermoelectric materials are essential to enhance the power factor while maintaining a low thermal conductivity. Preparation of such materials is the challenge, so far. It is possible now with $TiO_2$ doping. The influence of hierarchical $TiO_2$ inclusions on electrical and thermal transport properties has been investigated. Ba-filled skutterudite compound ($Ba_{0.3}Co_4Sb_{12}$) is synthesized by heating, annealing and ball milling. In this process, the prepared ingots are grinded in ball mill at 300 rpm for 6–8 hours to form the $Ba_{0.3}Co_4Sb_{12}$ powder matrix. Hierarchical $TiO_2$ nano-crystallite aggregates are prepared using a carbon sphere-template method and then dispersed into the $Ba_{0.3}Co_4Sb_{12}$ powder matrix at 300 rpm for 15 minutes using ball mill. Likewise, another commercial $TiO_2$ nanoparticles (consist of 30 % rutile and 70 % anatase) are dispersed in the powder matrix separately. It is found that $TiO_2$ inclusions enhance the Seebeck coefficient strongly and degrade the electric conductivity weakly. Also, it reduces the lattice thermal conductivity significantly [31].

### 6.4. $TiO_2$ – Industrial -Applications

$La_2O_3$ doped with $TiO_2$ nano-crystalline powder alloy is prepared to design improved electrodes for the energy storage applications. This $La_2O_3/TiO_2$ nano-alloy is prepared by mixing of both these powders using ball milling technique. Two varieties of samples are prepared *i.e.* sintered process and green process (without sintering). The degradation and changes in microstructure of these powders are analyzed. Electrochemical noise is utilized to analyze the influence of $La_2O_3$ with $TiO_2$. Both the prepared samples generate good response under NaOH solution. $La_2O_3/TiO_2$ nano-alloy electrode material shows good resistant at corrosion environment [32].

Alkaline earth metal titanates such as barium titanate ($BaTiO_3$) and strontium titanate ($SrTiO_3$) have some important properties. They are widely employed in the ceramic and electronic industries. $SrTiO_3$ is a para-electric, cubic structured, perovskite material at room temperature. It has various physical properties like large dielectric constant (~ 300), ferro-electricity, thermo-electricity, thermal conductivity (12 $Wm^{-1}K^{-1}$), photo-catalysis and low temperature (at < 20 K) superconductivity [33].

## 6.5. TiO$_2$ – Ceramic Process

Widodo *et al.* report about the milling process for the production of SrTiO$_3$ powder [34]. It is explicated in this sub-section. SrTiO$_3$ ceramic is prepared by mechanical alloying processing technique. Strontium carbonate (SrCO$_3$) and TiO$_2$ powders are mixed in vibrator ball mill followed by and heat treatment process. The milling process is done for 60 hours results in decreasing of mean particle size ~ 2 µm and crystallite size 17 nm and 19 nm respectively. The ball to powder ratio is 10:1 during milling.

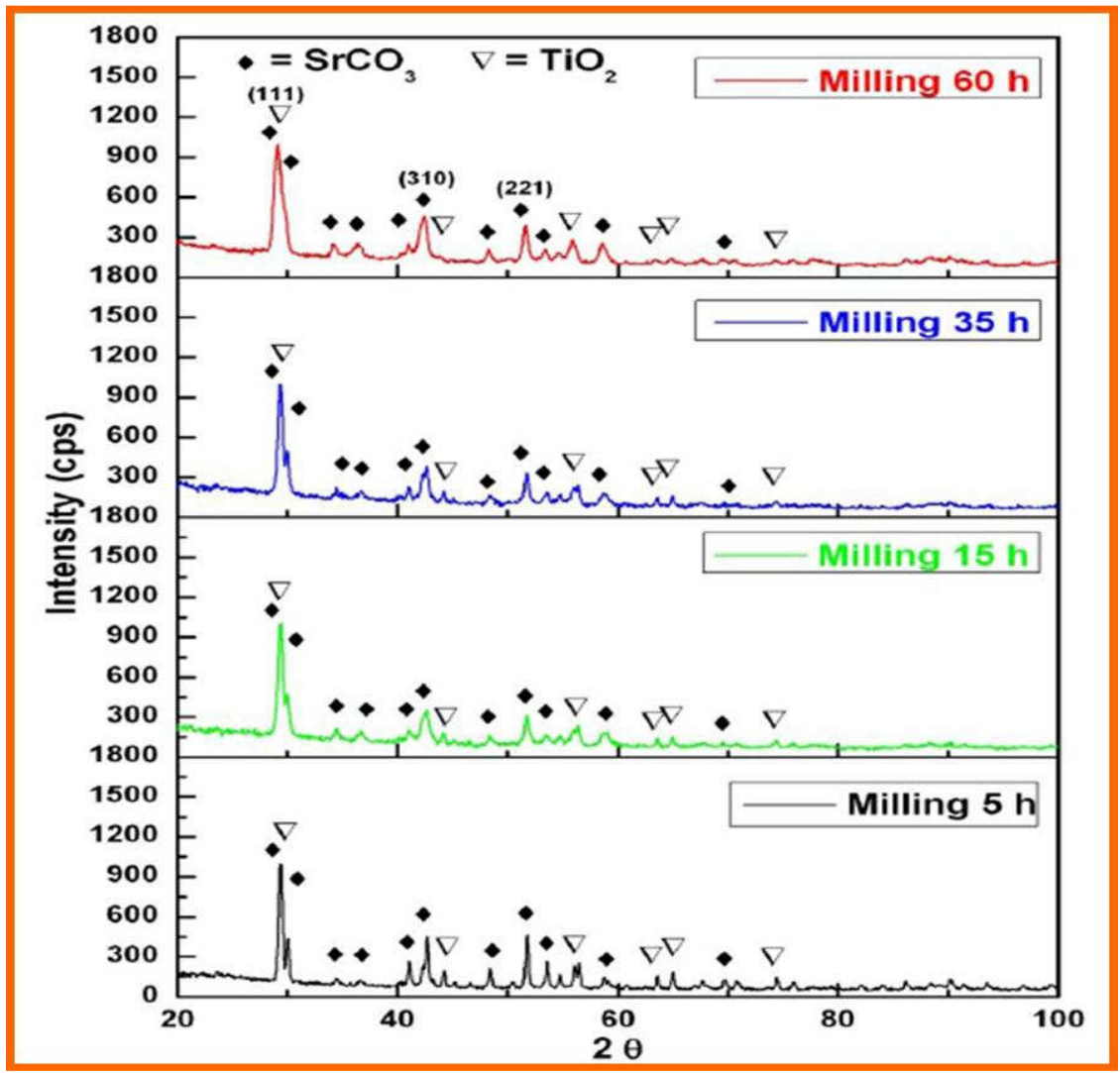

**Fig.7.** XRD of SrTiO$_3$ and TiO$_2$ upto 60 h milling. Fig. is adapted from [34].

Crystallite and microstructure of SrTiO$_3$ are required for specific application. Dense ceramic material is obtained by annealing at different temperatures. XRD of SrTiO$_3$ and TiO$_2$ powder after different time periods of milling are shown in Fig.7. XRD of the milled SrTiO$_3$ and TiO$_2$ powder after annealing at different temperatures (550 °C, 750 °C and 900 °C) for the duration of 0 h and 24 h are shown in Fig.8 to compare the effects of milling and annealing. It is revealed that heating of the particles at 900 °C temperature for 24 hours promotes the formation of SrTiO$_3$ nanocrystallites.

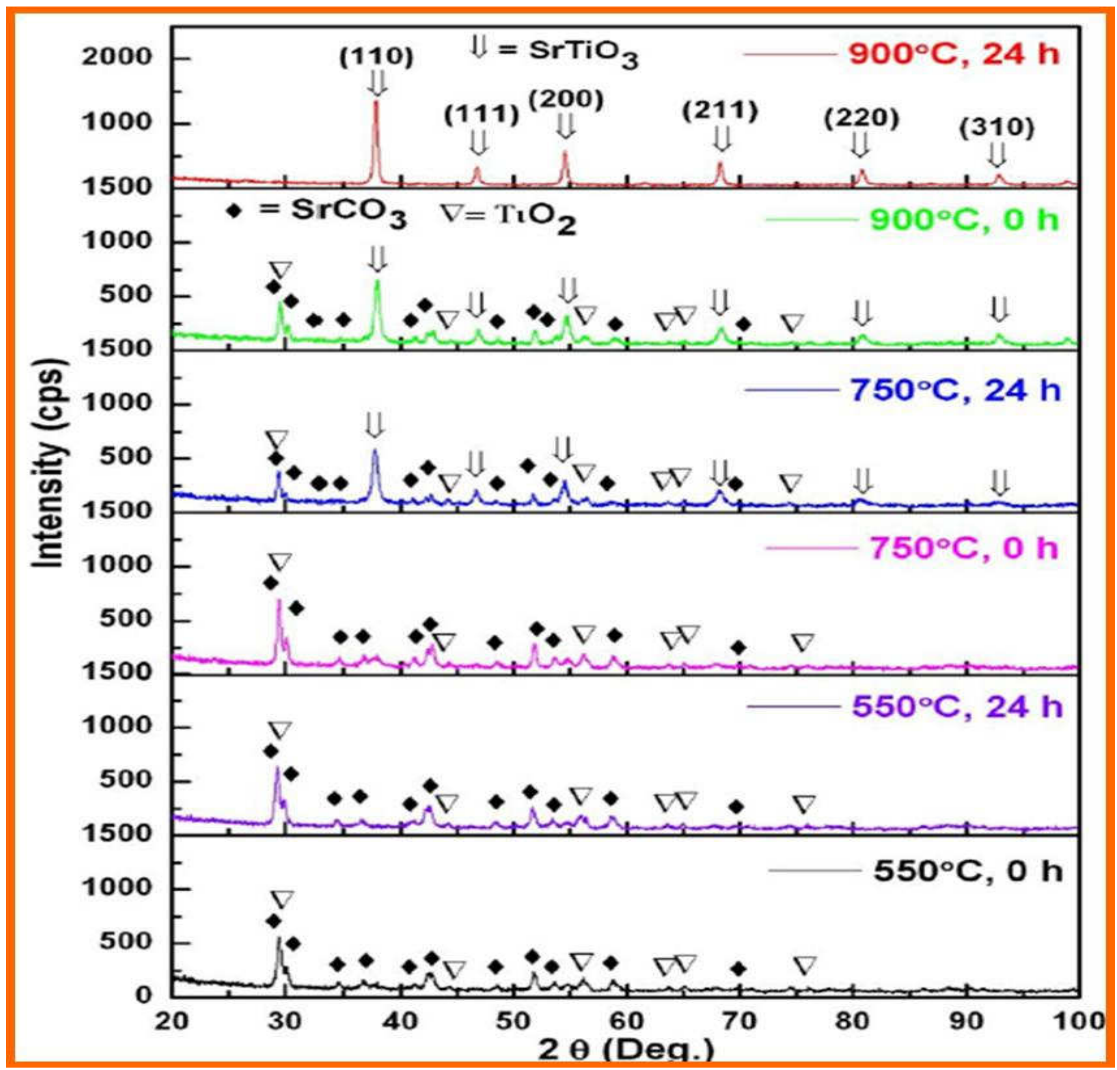

**Fig.8**. XRD of SrTiO$_3$ and TiO$_2$ up to 900 °C annealing. Fig. is adapted from [34].

## Conclusion

It is concluded from this review work that ball milling is used to produce nanomaterials in large scale. A high energy force is developed in ball mill during mechanical grinding by colliding of balls with the reactant media, and the cylindrical shell. It converts the bulk materials into nanomaterials. A repetitive fracturing and welding in this process lead to solid-state reaction resulting in the formation of nanomaterials. Milling of TiO$_2$ is done mainly for the purpose of size reduction and doping. Improving the efficiencies of TiO$_2$ such as photo-catalytic activity, mechanical strength and reducing the properties such as particle size, band gap, recombination rate of electron-hole pair are done using ball milling method. Rotating speed, grinding time, balls to grinding materials ratio and concentrations of the materials determine properties of the grinded product. Grinding of TiO$_2$ alone or TiO$_2$ as doping material with other materials or other doping materials with TiO$_2$ creates new properties in these materials or creates new composites materials which are exploited in many applications.